\documentclass[a4paper,amsmath,amssymb]{jpconf}
\usepackage{graphicx}
\usepackage{amsmath, amsthm, amssymb}
\usepackage{multirow}
\begin{document}
\title{Explosive percolation in graphs}

\author{Santo Fortunato}

\address{Complex Networks and Systems Lagrange Laboratory, ISI
  Foundation, Viale S. Severo 65, 10133, Torino, Italy}

\ead{fortunato@isi.it}

\author{Filippo Radicchi}

\address{Chemical and Biological Engineering,
Northwestern University, Evanston, IL 60208, USA}

\ead{f.radicchi@gmail.com}

\begin{abstract}
Percolation is perhaps the simplest example of a process exhibiting a phase
transition and one of the most studied phenomena in statistical
physics. The percolation transition is continuous if sites/bonds are
occupied independently with the same probability. However, alternative
rules for the occupation of sites/bonds might affect the order of the 
transition. A recent set of rules proposed by Achlioptas et al. [{\it
  Science} {\bf 323}, 1453 (2009)], characterized by competitive link
addition, was claimed to lead to a discontinuous connectedness
transition, named ``explosive percolation''. In this work we survey a
numerical study of the explosive percolation transition on various
types of graphs, from lattices to scale-free networks, and show the
consistency of these results with recent analytical work showing that
the transition is actually continuous.

\end{abstract}

\section{Introduction}
\label{sec:Introduction}

Phase transitions lie at the heart of the modern development of
statistical physics~\cite{stanley71,binney92}. They are changes in the
state of order of a system and can 
be classified based on their properties in the immediate proximity of
the critical point. Phase transitions are
{\it continuous} (or second-order), if the order parameter
changes continuously across the two phases, with an infinite
correlation length and consequent power law decay of correlations and
divergence of higher moments of the order parameter at the critical point.
Otherwise one speaks of {\it discontinuous} (first-order) phase
transitions, which are typically characterized by a discontinuous jump
of the order parameter at the critical point. 

The purely geometric process known as {\it random
  percolation}~\cite{staufferbook} offers a paradigmatic example of
a continuous phase transition. The starting point is a graph, e. g. a
lattice. The sites or the links of the lattice are occupied
independently with some probability $p$. Nearest-neighboring occupied
sites/links form structures called {\it clusters}. For low values of the
occupation probability $p$ just a few small clusters are formed, but
if $p$ increases the number and size of the clusters will increase as
well. When the occupation probability exceeds a critical value $p_c$,
a macroscopic cluster, occupying a finite fraction of the total number
of sites/links, emerges. This macroscopic structure is called
{\it percolation cluster} and its relative size $P$, the {\it percolation
strength}, is the order parameter of
the transition: $P=0$ indicates the phase with only microscopic
clusters, whereas $P>0$ indicates the phase with (at least) one 
macroscopic cluster. Random percolation has been studied on lattices, random
graphs~\cite{erdos} and scale-free networks~\cite{cohen00, newman01,
  pastor00, dorogovtsev08, vazquez04}. Analytical and numerical
studies have proved that the percolation transition is continuous,
without exceptions. This however holds for random percolation. It
cannot be excluded {\it a priori} that alternative processes of
occupation of sites/links might lead to different types of geometric
transitions. 

\begin{figure}
\begin{center}
\includegraphics[angle=-90,width=10cm]{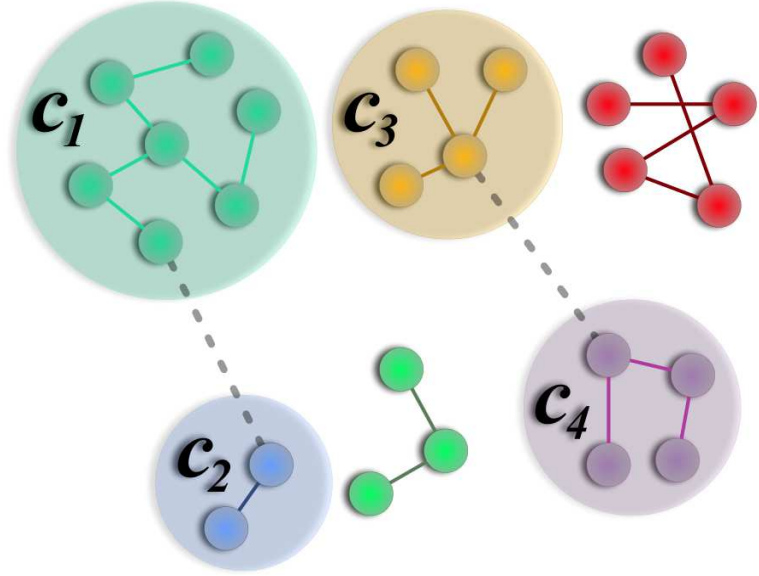}
\end{center}
\caption{Scheme of an Achlioptas process with product rule. Two links (dashed lines) are
selected at random among all possible pairs of non-adjacent
nodes. The link joining the pair of clusters with the smaller product
size is eventually added to the system. Here
the winning link is that between clusters
$c_1$ and $c_2$ (product size $7\cdot 2=14<4\cdot 4=16$).}
\label{fig1}
\end{figure}

In a recent paper~\cite{achlioptas09}, Achlioptas et al. have
introduced a special set of rules, in which links are occupied as a result of a competitive
process between pairs of links ({\it Achlioptas processes}). The idea is to slow down the process
of cluster growth, by inserting links leading to the merge of small
clusters. This can be done in several ways. Achlioptas et al. focused
on the so-called {\it product rule} (PR): given a pair of links,
randomly selected among those which are not yet occupied, one occupies
the link merging the two clusters with smaller product size
(Fig.~\ref{fig1}). One could consider variants of this rule, like
taking the sum instead of the product, or just the minimum size of the
clusters of each pair. Also, the competition can be extended to more
than two links. In any case, the result of such processes is a slow
growth of the cluster sizes, which causes a delay in the onset of the
percolation transition. On the other hand, since the density of links
and, consequently, of clusters
at the onset is higher than for random percolation, it is natural to
expect that the percolation cluster has a very rapid
growth. This is indeed confirmed by numerical studies; in fact, the
growth of the percolation cluster is so quick that the percolation
strength $P$ appears to vary discontinuously at the onset. The sudden
jump in the order parameter has motivated the name ``explosive
percolation''. In the last two years, Achlioptas
processes have been extensively studied and meanwhile a lot is known about
the explosive percolation
transition~\cite{ziff09,cho09,radicchi09,friedman09,radicchi10,moreira10,souza10,ziff10,cho10,araujo10,nagler10,dacosta10}.  

The main issue concerned the order of the transition: is it continuous
or discontinuous? Achlioptas et al. claimed that it is discontinuous,
and this is what has mostly attracted the attention of scholars. As a
matter of fact, it was soon shown that, despite the alleged jump  
of the order parameter at the threshold $p_c$, the explosive percolation
transition has peculiar features of continuous phase transitions, like
power law distributions of cluster sizes~\cite{radicchi10,souza10,cho10} and power law scaling of the
mean cluster size at $p_c$~\cite{cho09,radicchi09,radicchi10}. Indeed, in recent
works by Nagler et al.~\cite{nagler10} and da Costa et
al.~\cite{dacosta10} it was proven that the transition is actually
continuous, and that the jump of the order parameter is only apparent,
and due to the very small critical exponent of the order parameter $\beta$.

In this paper we summarize the finite size scaling analysis of
Ref.~\cite{radicchi10}, and verify that the results are indeed
consistent with the analytical findings of da Costa et al.. In
Section~\ref{sec1} we introduce finite size scaling, in
Section~\ref{sec2} we discuss the results for different types of
graphs. A summary is reported in Section~\ref{sec3}.

\section{Finite size scaling}
\label{sec1}

Finite size scaling~\cite{landau00} is a well-known technique adopted
in numerical studies of phase transitions. For continuous phase
transitions, the correlation length is infinite at the critical
threshold $p_c$, so every variable $X$ 
is scale-independent in the proximity of that point and has a power law form,
\begin{equation}
X \sim |p-p_c|^{\omega},
\label{eqscal1}
\end{equation}
where $\omega$ is a critical exponent. If the system has a finite size
$N$, the variable $X$ near the threshold has the following scaling form 
\begin{equation}
X = N^{-\omega/\nu}F\left[ \left(p-p_c\right)\, N^{1/\nu} \right].
\label{eqscal2}
\end{equation}
In Eq.~\ref{eqscal2}, $\nu$ is a critical exponent and $F$ a universal function.
Exactly at the critical point ($p=p_c$) the variable displays the simple scaling $X \sim N^{-\omega/\nu}$, which can be used to extract the exponents' ratio
$\omega/\nu$, by using systems of different sizes. Moreover, if $p_c$, $\nu$ and $\omega$ are known,
the expression $XN^{\omega/\nu}$ as a function of
$\left(p-p_c\right)\, N^{1/\nu}$ is just the universal 
function $F$, which is independent of $N$, so curves referring to different system sizes collapse.

We investigated the two main variables of percolation~\cite{staufferbook}, i.e. the {\it percolation strength} $P$
and the {\it average cluster size} $S$. The percolation strength $P$,
as we have said above, is the order parameter of the transition, and equals
the relative size of the percolating cluster(s) with respect to the
total system size $N$. While on lattices there are operative criteria
to define a percolating cluster (e.g. if it runs from one edge of the
lattice to the opposite one, say), on generic graphs this is not the case,
so $P$ is defined as the relative size
of the largest connected cluster. The scaling ansatz of the
percolation strength is 
\begin{equation}
P = N^{-\beta/\nu}F^{(1)}\left[ \left(p-p_c\right)\, N^{1/\nu} \right].
\label{eqP}
\end{equation}
The average cluster size $S$ is defined as 
\begin{equation}
S = \frac{\sum_{s} n_s s^2}{\sum_{s}n_s s} \;,
\end{equation}
where $n_s$ is the number of clusters of size $s$ per node.
The sums run over all possible values of
$s$ except for the one of the largest cluster.
The scaling ansatz of $S$ is 
\begin{equation}
S = N^{\gamma/\nu}F^{(2)}\left[ \left(p-p_c\right)\, N^{1/\nu} \right].
\label{eqS}
\end{equation}
The universal functions $F^{(1)}$ and $F^{(2)}$ of Eqs.~(\ref{eqP}) and~(\ref{eqS}) 
are not the same, but they are related.

In random percolation, the probability distribution $P(s)$ of cluster
sizes (except the largest), 
decays at $p_c$ as the power law $P(s)\sim s^{-\tau}$ with the cluster size $s$.
We have computed the cluster size distribution $P(s)$ at $p_c$ and
measured the Fisher exponent $\tau$. 
For a given system, $P(s)$ is related to $n_s$ by
the relation $P(s)=Nn_s/n_c$, where $n_c$ is the total number of ``finite'' clusters.
We shall use the symbol $n_s$ to indicate $P(s)$ as well, but in the
plots $n_s$ is normalized as $P(s)$, 
for consistency. 

The percolation threshold $p_c$ is localized in two independent ways.
The first method exploits the scaling of the pseudo-critical points $p_c(N)$
\begin{equation}
p_c = p_c(N) + b N^{-1/\nu} \,.
\label{eq:chi2}
\end{equation}
By using several system sizes, one can perform a fit with the three parameters
$b$, $\nu$ and $p_c$. The pseudocritical point for a system with
finite size $N$ can be defined in
several ways, for us it indicates the value of $p$ at which the
average cluster size $S$ peaks.

An alternative procedure relies on Eq.~(\ref{eqP}). The percolation 
strength $P$ is plotted as a function of the system size $N$ for a
given value of $p$. Since, for $p=p_c$ the scaling follows a power law, the correct
value of the percolation threshold can be determined by finding the value of $p$ which yields the best power law fit.

\section{Results}
\label{sec2}

\subsection{Implementing Achlioptas processes with product rule}

The starting point is just a graph with $N$ nodes and no links. Links
are added one by one, according to the competitive rule previously
described, i.e. by selecting each time a pair of links at random and picking the
one yielding the smaller product for the sizes of the clusters it
merges.
For scale-free networks the situation is a bit more involved, and we
describe it in Section~\ref{scf}. 
The procedure goes on until
the desired density of links $p$ is reached. We defined $p$ as the number of links of the graph divided by 
the total number of links present in the graph when it has been
``completed'', i.e., after the addition of the last link.
All graphs considered here are ``sparse'', i.e., the average degree
$\langle k\rangle$, expressing the ratio between (twice) 
the number of links and the number of nodes $N$,
does not depend on $N$. At time $t$ of the growth process there are exactly $t$ links in the system:
their density $p$, according to our definition, is then $t/(N{\langle k\rangle})$.  

\subsection{Lattices}
\label{lat}

\begin{figure}[htb]
\begin{center}
\includegraphics[width=0.32\textwidth]{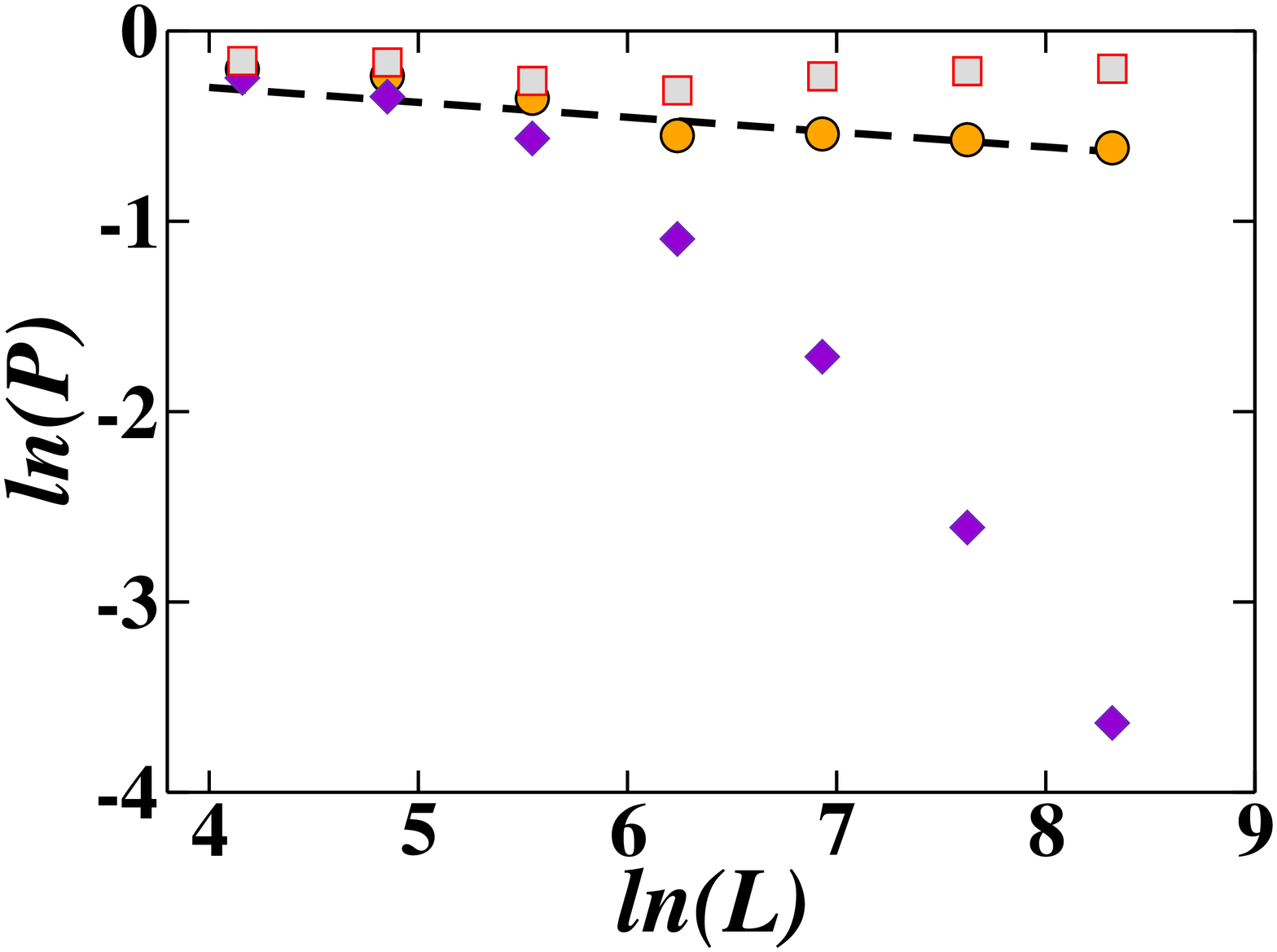}
\includegraphics[width=0.32\textwidth]{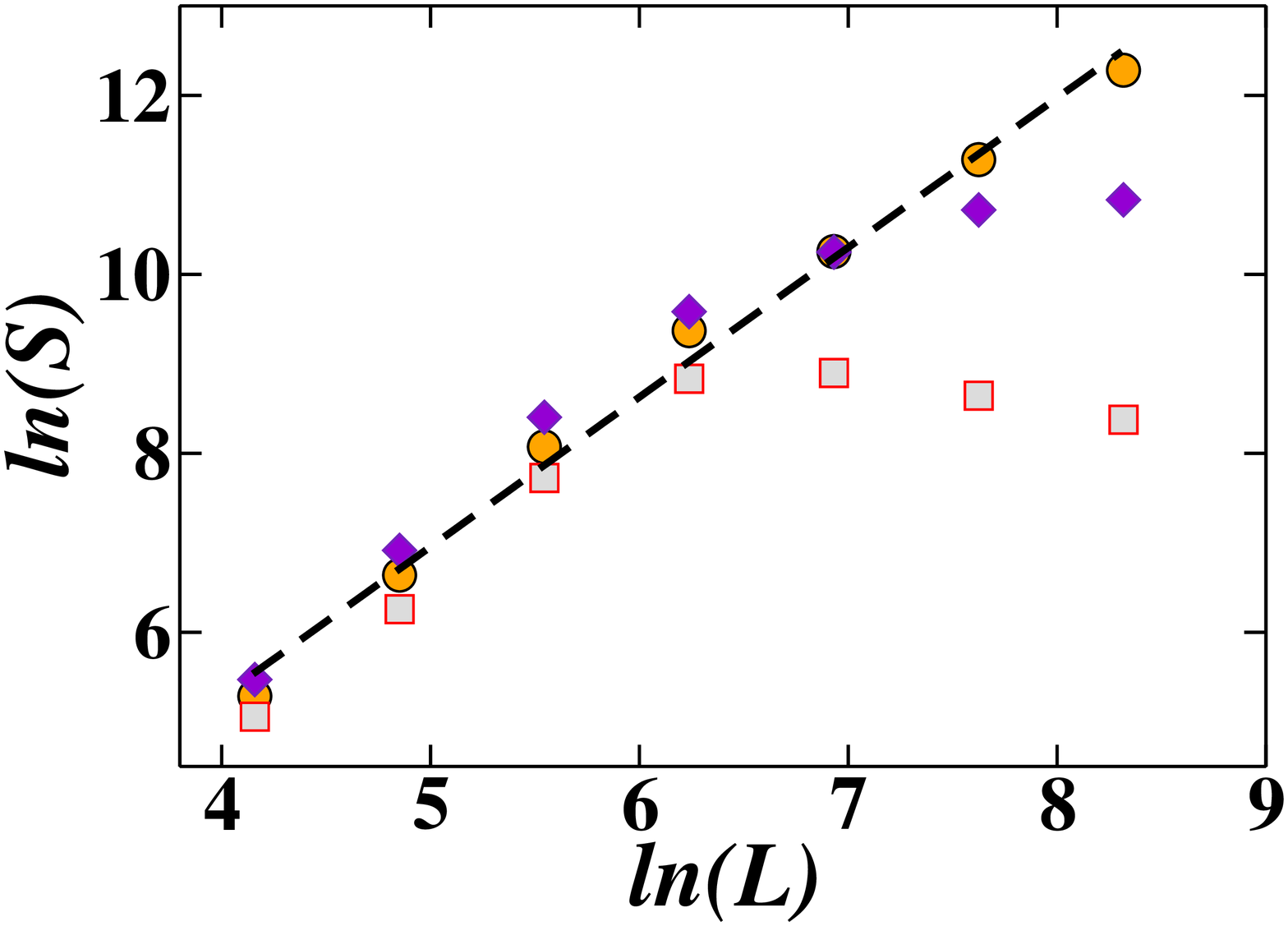}
\includegraphics[width=0.32\textwidth]{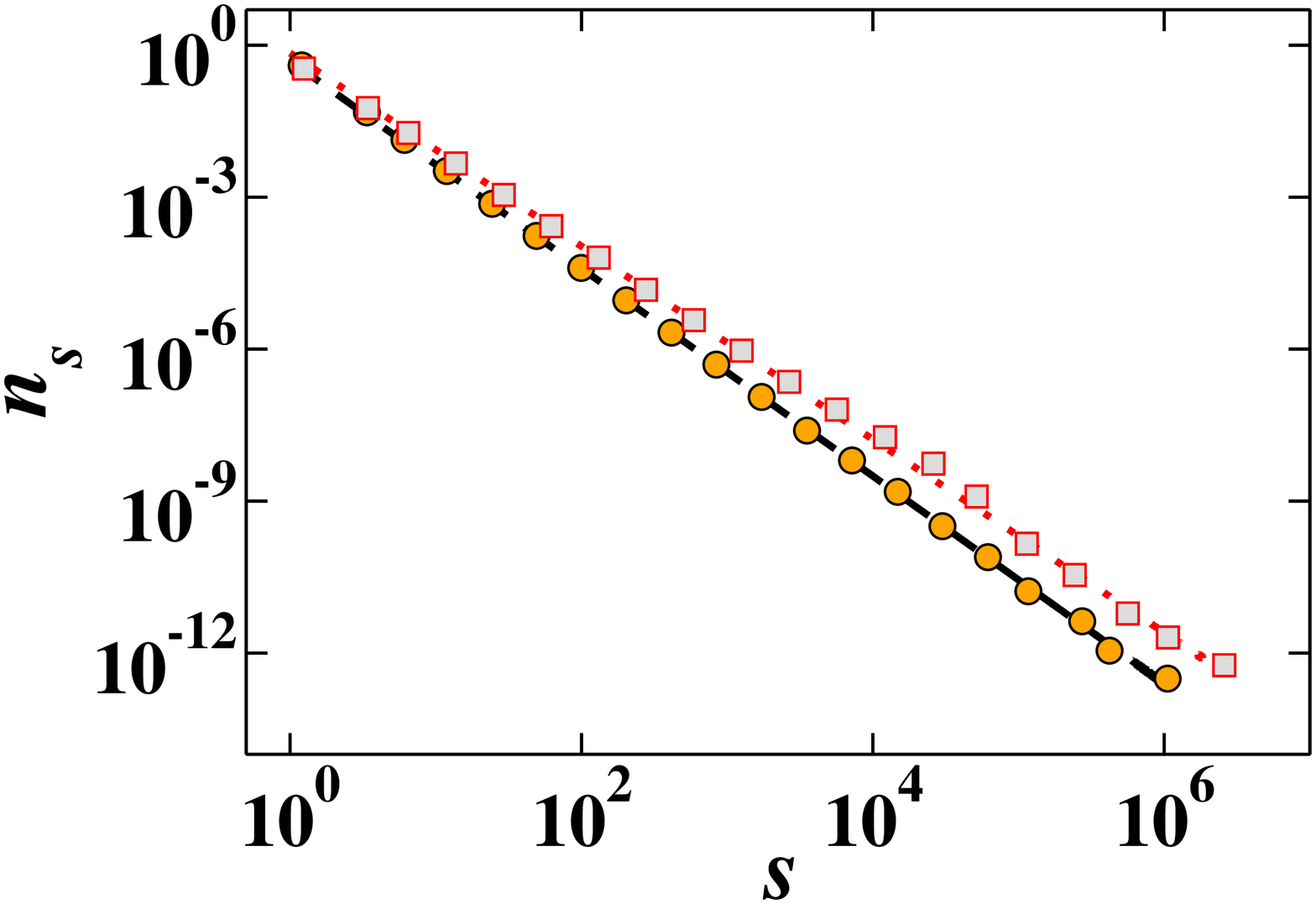}
\caption{Explosive percolation on $2d$-lattices. 
(Left) Percolation strength $P$ as a function of the lattice side $L$ for three different 
values of the occupation probability: $p=0.5256$ (violet diamonds), $p=0.5266$ (orange circles) and $p=0.5276$ 
(grey squares). The dashed line indicates the best fit obtained at the
critical point $p=p_c=0.5266(2)$, from which we get 
$\beta/\nu=0.07(3)$. (Center) The average cluster size $S$ as a function of the lattice side $L$ for the same 
values of $p$ used for $P$. The dashed line has slope
$\gamma/\nu=1.7(1)$. (Right) Cluster 
size distributions measured at $p_c$ for PR (grey squares) and random
percolation (orange circles). The exponents are $\tau=1.9(1)$ for PR
(red dotted line), while for random percolation $\tau=2.05(1)$ (black dashed line). Simulations have been 
performed on lattices with side $L=4096$.}
\label{fig2}
\end{center}
\end{figure}

Achlioptas processes on the square lattice were first studied by
Ziff~\cite{ziff09}, who found similar properties for the
explosive transition as Achlioptas et al. had found for
Erd\"os-R\'enyi random graphs. The criterion to assess the nature of
the transition was the same one proposed by Achlioptas et al., namely
the scaling with the system size $N$ of the transition window $\Delta
p=p_2-p_1$, where $p_2$ is the lowest value of $p$ for which $P > 0.5$ and 
$p_1$ the lowest value of $p$ for which $P > 1/\sqrt{N}$. In a recent
paper Ziff has performed a finite size scaling analysis as well~\cite{ziff10}.
The results of our analysis are shown in Fig.~\ref{fig2}.
The order parameter $P$ yields a very small exponents' ratio $\beta/\nu$,
compatible with zero [$0.07(3)$] (Fig.~\ref{fig2}, left), which is what one
would expect to find for a discontinuous transition. 
The average cluster size $S$, instead, has a non-trivial power law scaling at $p_c$, with exponent
$\gamma/\nu=1.7(1)$ (Fig.~\ref{fig2}, center). This 
had first been observed by Cho {\it et al.} in scale-free networks~\cite{cho09}. 
Fig.~\ref{fig2} (right), showing
the distribution of sizes $n_s$ for all clusters except the largest
one, provides an explanation of the scaling of $S$. The distribution is a clear power law 
[exponent $1.9(1)$], which is incompatible with 
a classic discontinuous transition, as it usually occurs for continuous transitions. 
Since $n_s$ is a power law, all variables derived from $n_s$, including the average
cluster size $S$, have power law scaling.

\begin{figure}
\begin{center}
\includegraphics[width=0.32\textwidth]{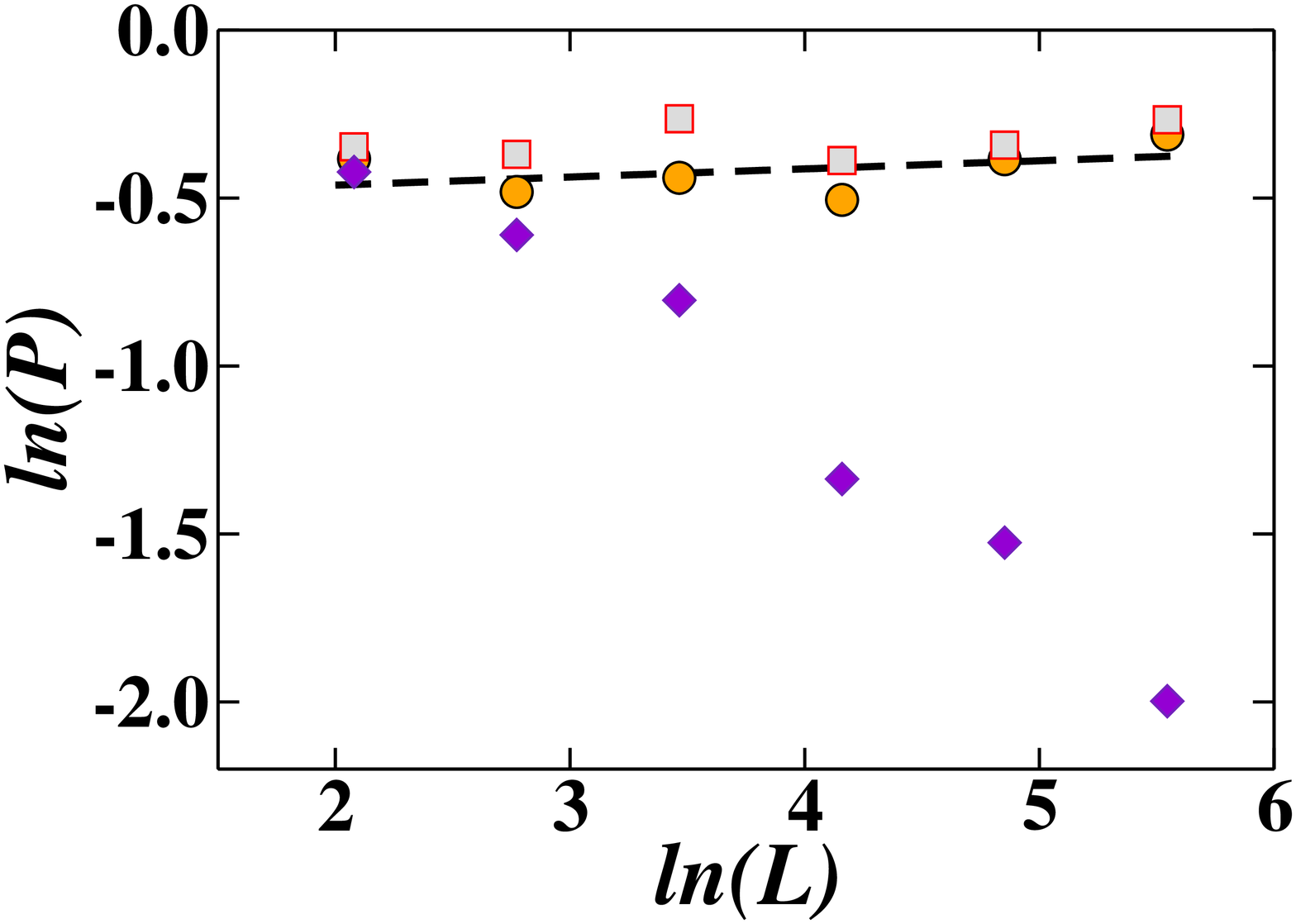}
\includegraphics[width=0.32\textwidth]{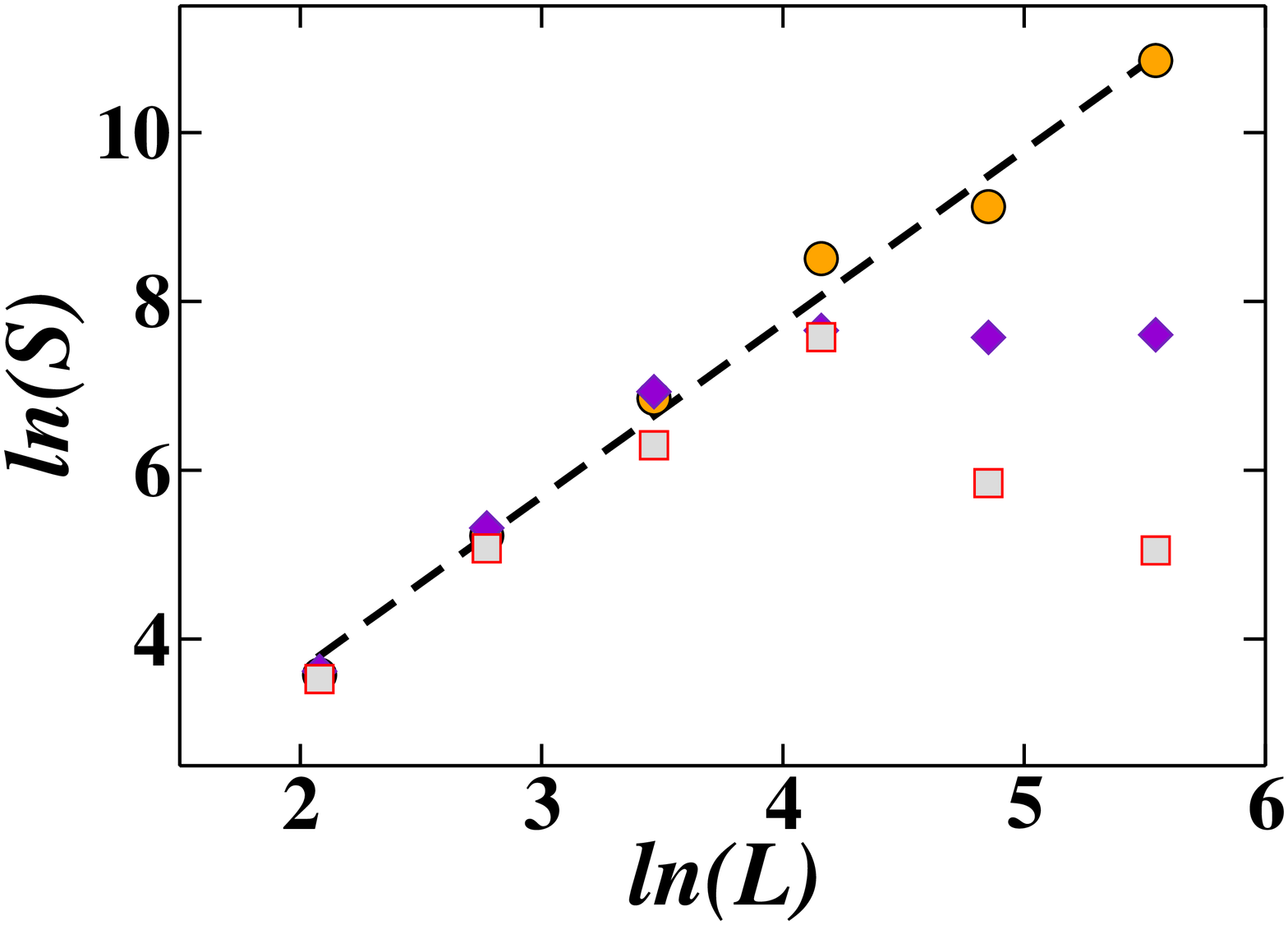}
\includegraphics[width=0.32\textwidth]{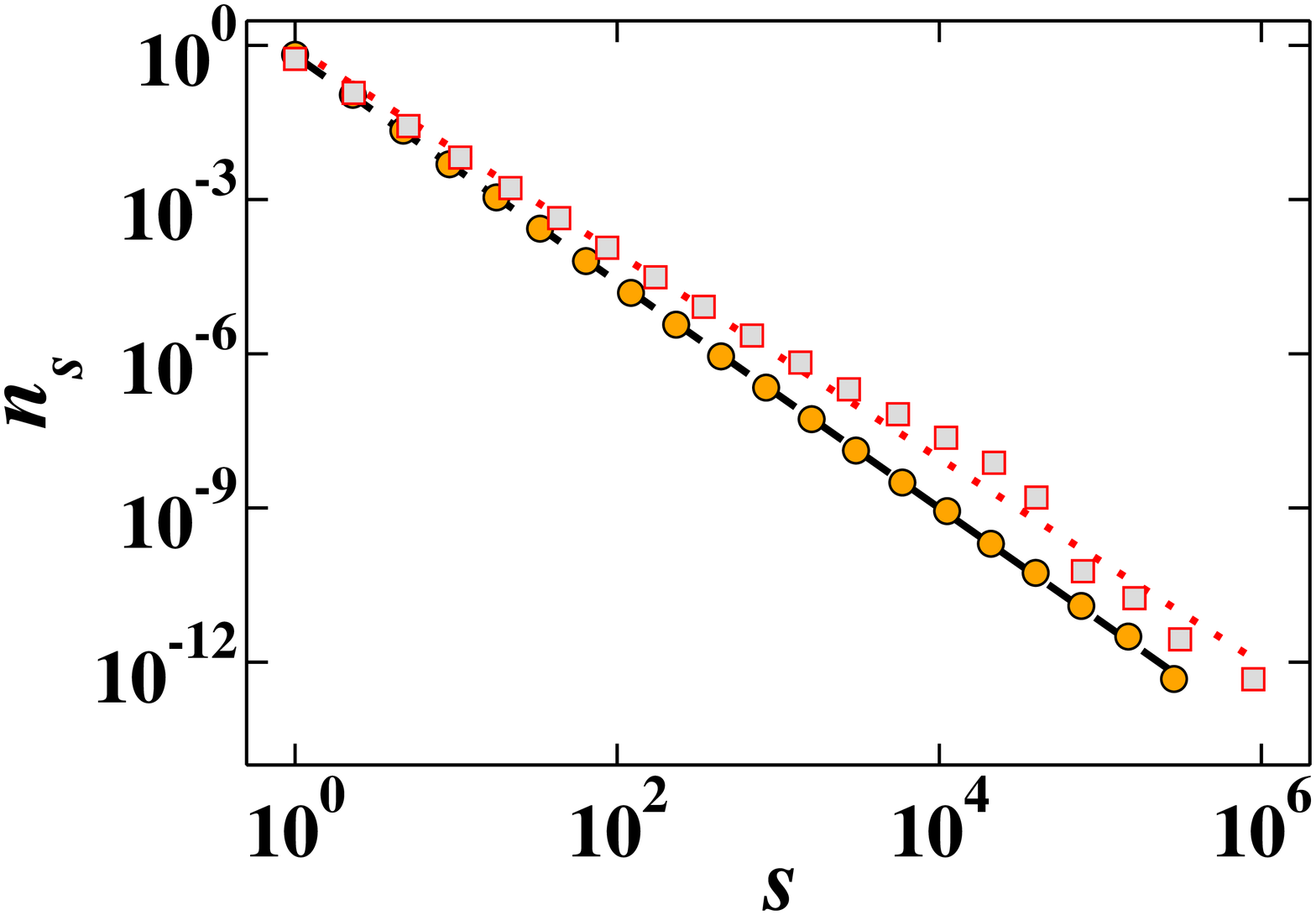}
\caption{Explosive percolation on $3d$-lattices. (Left) Percolation strength $P$ as a 
function of the lattice side $L$ for three different values of the occupation probability: $p=0.3866$ 
(violet diamonds), $p=0.3876$ (orange circles) and $p=0.3886$ (grey squares). The dashed line indicates 
the best fit obtained at the critical point $p=p_c=0.3876(2)$, from
which we get $\beta/\nu=0.02(2)$. 
(Center) The average cluster size $S$ as a function of the lattice side $L$ for the same 
values of $p$ used for $P$. The dashed line has slope $\gamma/\nu=2.1(1)$. (Right) Cluster size distributions 
at the critical point for PR (grey squares) and random percolation
(orange circles). The exponents of the power law fits are
$\tau=1.99(4)$ for PR (red dotted line) and $\tau=2.20(1)$ for random
percolation (black dashed line). Simulations 
have been performed on lattices with side $L=256$.}
\label{fig3}
\end{center}
\end{figure}

In 3d-lattices, the general picture is consistent with that in two dimensions (Fig.~\ref{fig3}). 
Again, the scaling at $p_c$ of the order parameter $P$ yields a very
small exponent $\beta/\nu=0.02(2)$, compatible with zero
(Fig.~\ref{fig3}, left). 
Still, $S$ scales with an exponent $\gamma/\nu=2.1(1)$
(Fig.~\ref{fig3}, center), again due to the power law shape 
of the distribution of cluster sizes (Fig.~\ref{fig3}, right). The
Fisher exponent $\tau=1.99(4)$ is compatible with
the value we measured in two dimensions [$1.9(1)$].

\subsection{Erd\"os-R\'enyi graphs}
\label{sec:er}

\begin{figure}
\begin{center}
\includegraphics[width=0.32\textwidth]{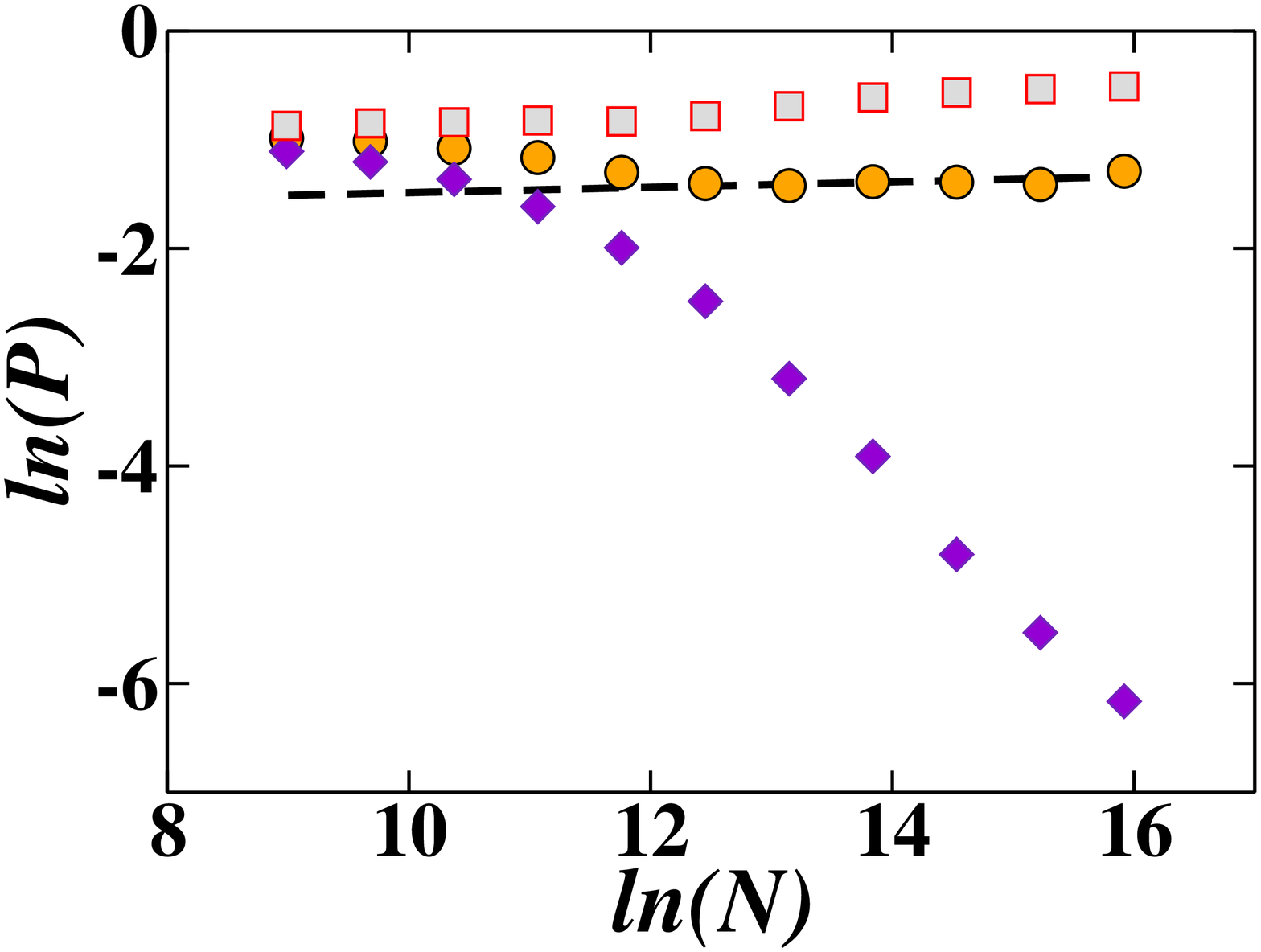}
\includegraphics[width=0.32\textwidth]{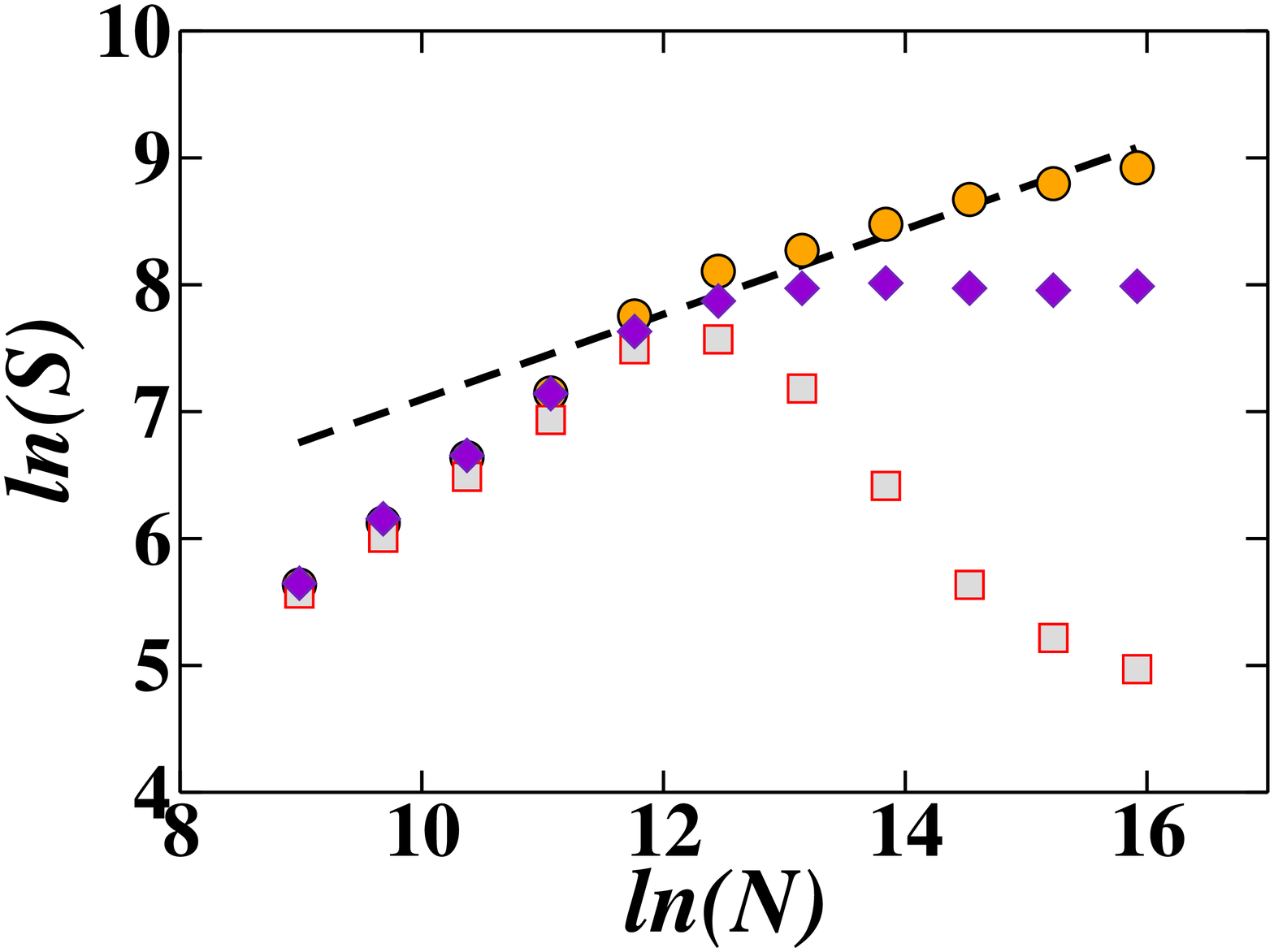}
\includegraphics[width=0.32\textwidth]{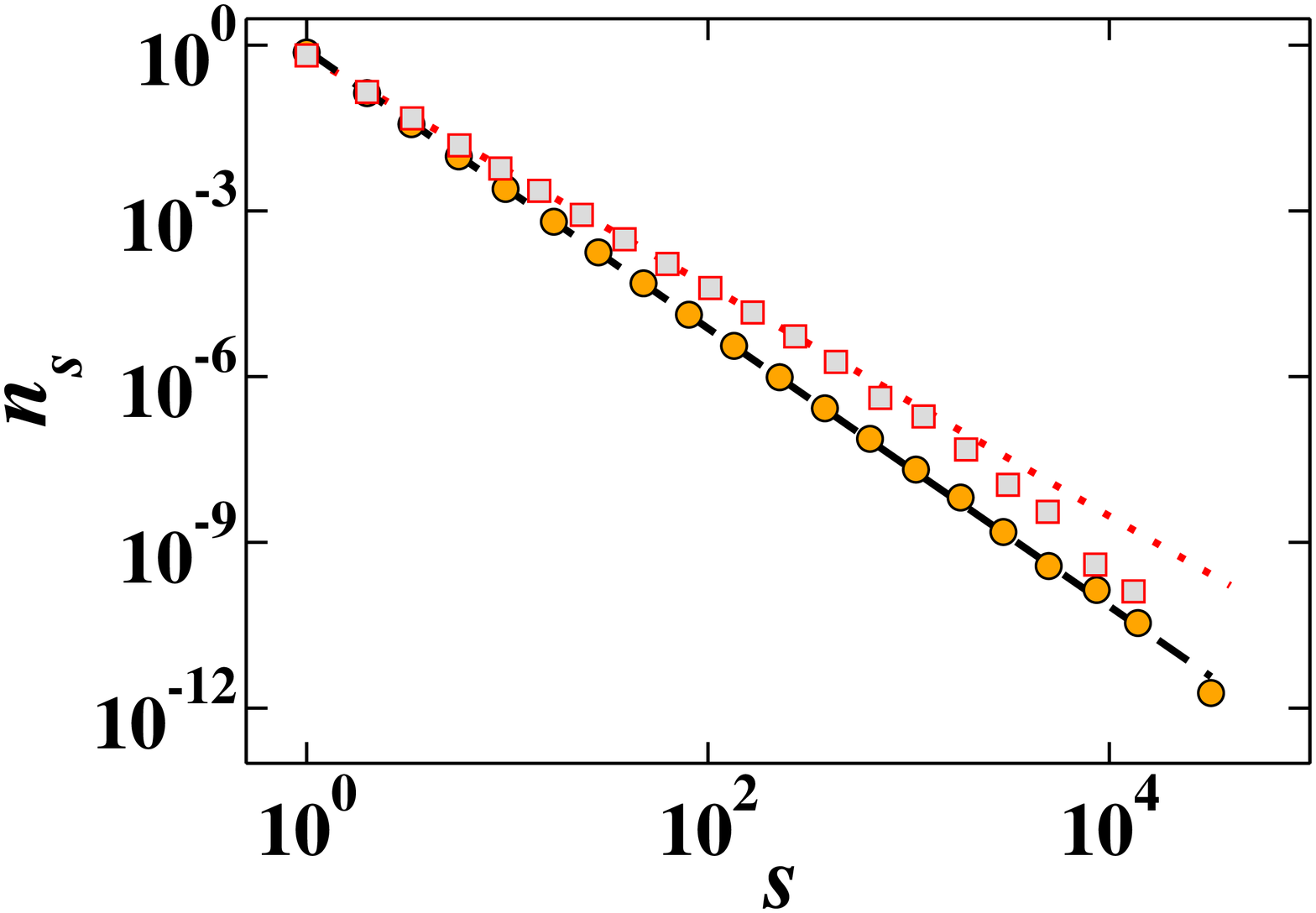}
\caption{Explosive percolation on Erd\"os-R\'enyi random graphs. (Left) Percolation strength $P$ as a function of the 
network size $N$ for three different values of the occupation probability: $p=0.8872$ (violet diamonds), 
$p=0.8882$ (orange circles) and $p=0.8892$ (grey squares). The dashed line indicates the best fit obtained 
at the critical point $p=p_c=0.8882(2)$, from which we get $\beta/\nu=0.02(1)$. (Center)
The average cluster size $S$ as a function of the network size $N$ for
the same values of $p$ used for $P$. 
The dashed line has slope $\gamma/\nu=0.48(4)$. (Right) Cluster size
distributions $n_s$ measured at the critical point for PR (grey
squares) and random percolation (orange circles). The exponents are 
$\tau=2.08(5)$ (red dotted line) for PR and $\tau=2.51(2)$ (black
dashed line) for random percolation.
Simulations have been performed on graphs with $N=8\,192\,000$.}
\label{fig4}
\end{center}
\end{figure}

Fig.~\ref{fig4} presents the results of our analysis. For PR
we see again a flat profile of the order parameter $P$ with $N$
($\beta/\nu=0.02(1)$), consistent with a discontinuous transition, along with a power law scaling of the average
cluster size $S$, with exponent $\gamma/\nu=0.48(4)$. The exponent
$\tau=2.08(5)$ (Fig.~\ref{fig4}, right) is still compatible with the values found for 
PR on lattices, both in two and three dimensions (see Section~\ref{lat} and
Table 1). 

\subsection{Scale-free networks}
\label{scf}

\begin{figure}
\begin{center}
\includegraphics[width=0.6\textwidth]{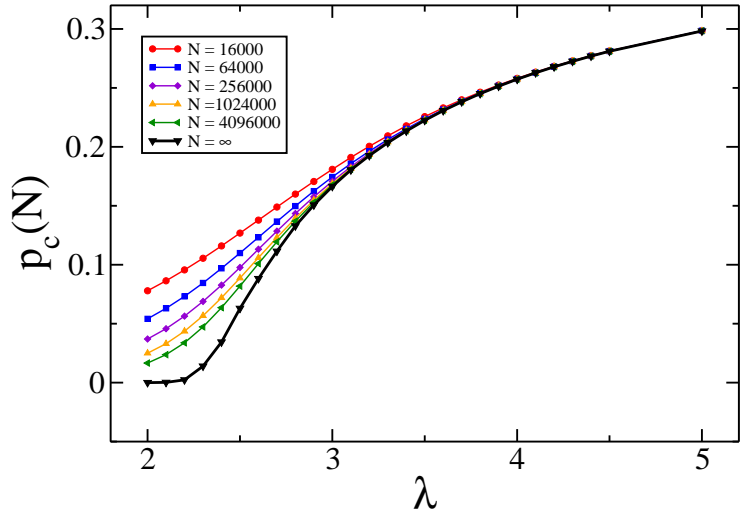}
\end{center}
\caption{Explosive percolation on random SF networks. Percolation
threshold $p_c(N)$ as a function of the degree exponent $\lambda$ for different network sizes $N$. 
The black line is the infinite size limit extrapolation of the critical threshold, performed by applying 
Eq.~(\ref{eq:chi2}). The percolation threshold vanishes for
$\lambda<\lambda_c\sim 2.3$, while it is non-zero for $\lambda>\lambda_c$.}
\label{figthres}
\end{figure}
\begin{figure}
\begin{center}
\includegraphics[width=0.45\textwidth]{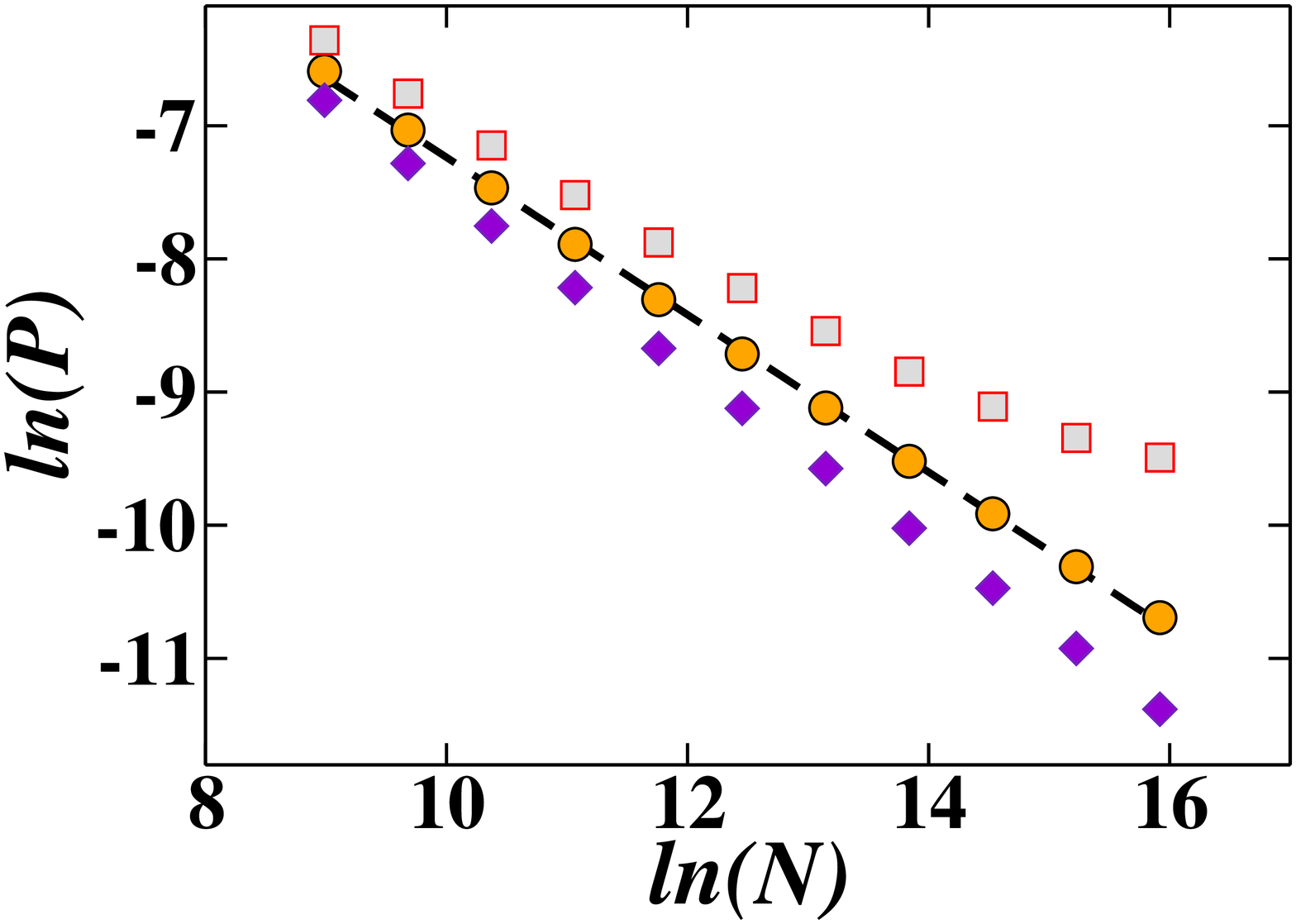}
\includegraphics[width=0.45\textwidth]{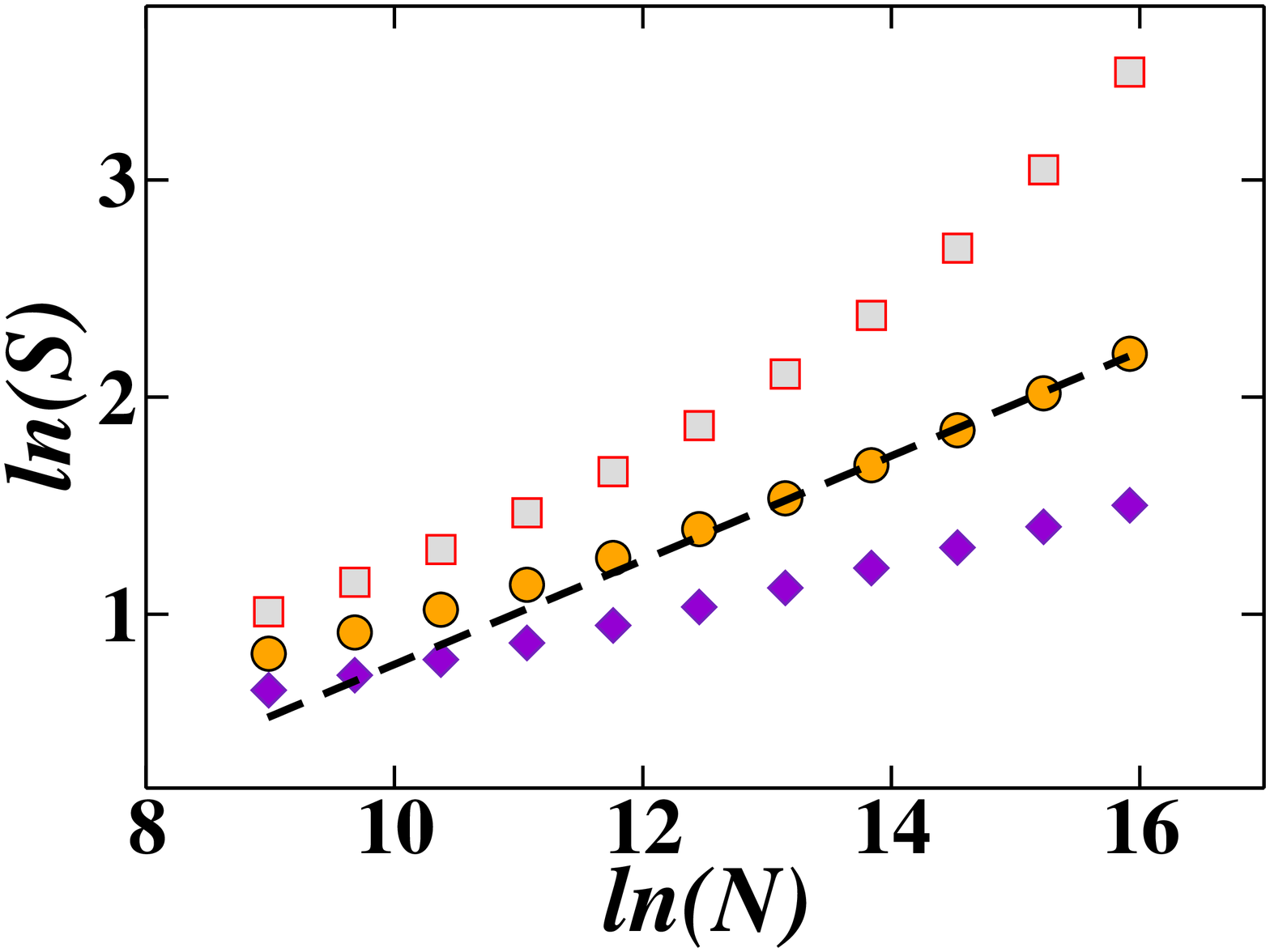}
\vskip1cm
\includegraphics[width=0.45\textwidth]{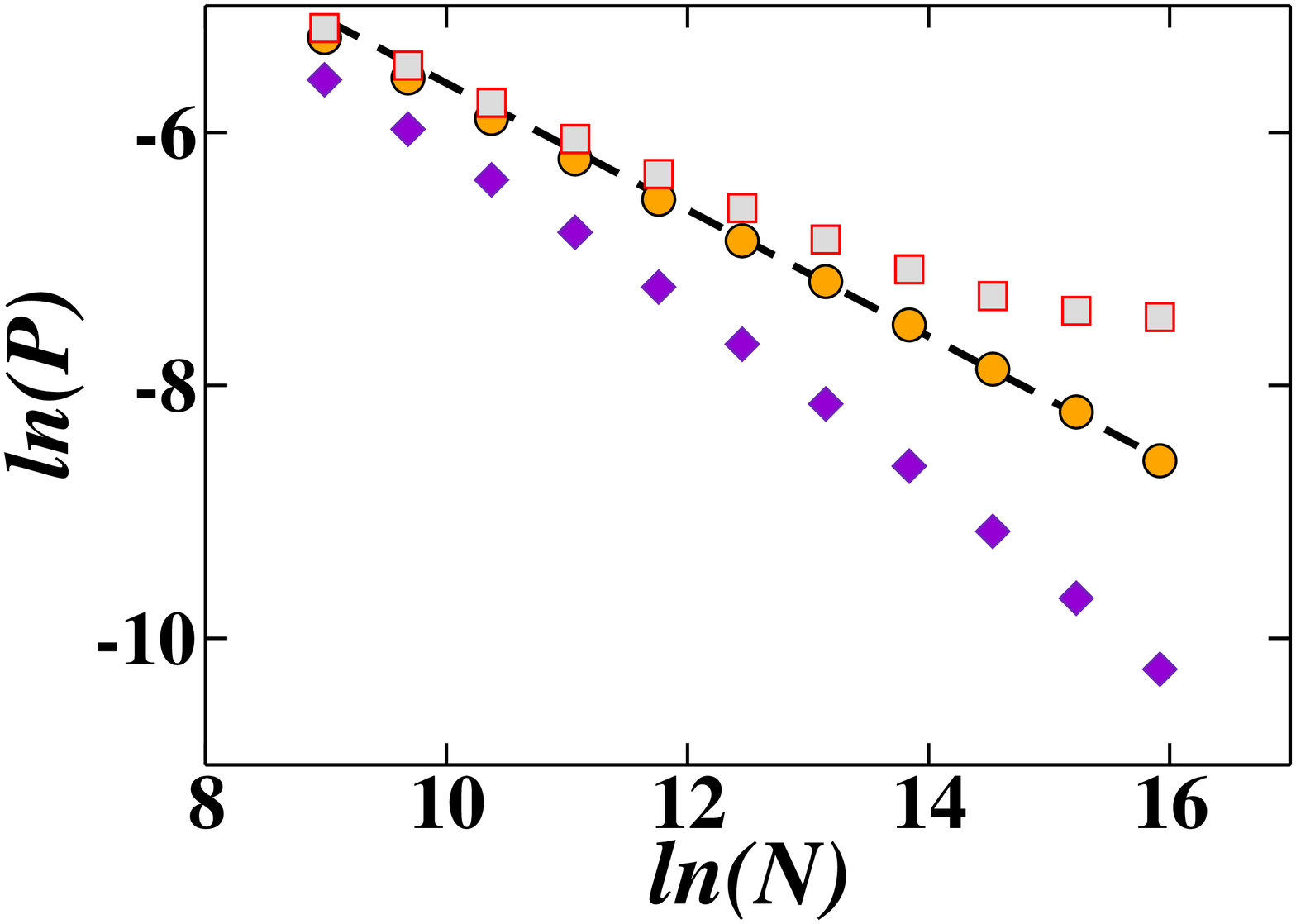}
\includegraphics[width=0.45\textwidth]{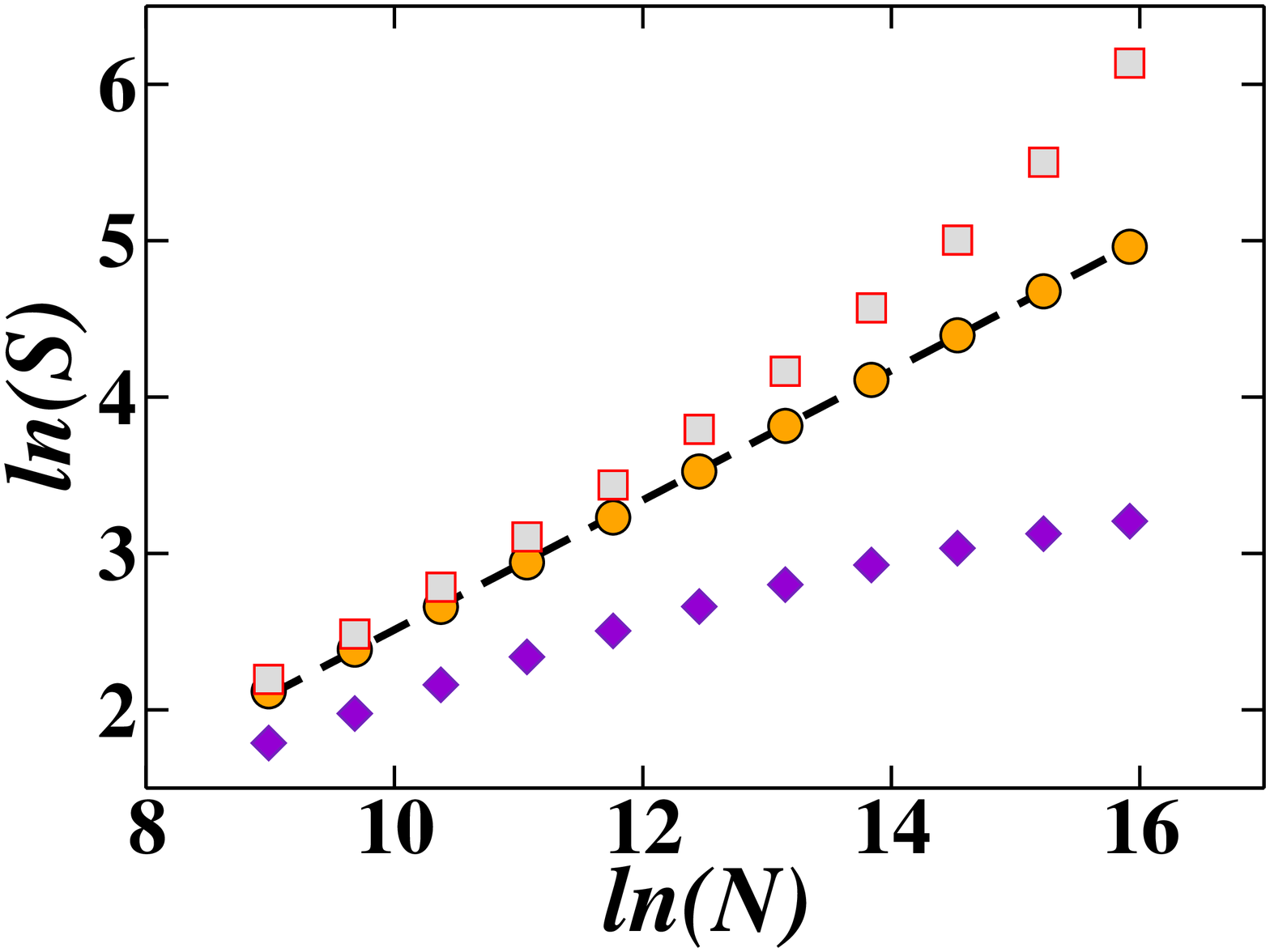}
\caption{Explosive percolation on SF networks for degree exponents
$\lambda=2.5$ (top diagrams) and $\lambda=2.8$ (bottom diagrams). 
(Left) Percolation strength $P$ as a function of the system size $N$ for three 
different values of the occupation probability. Top: $p=0.0529$ (violet diamonds), $p=0.0629$ 
(orange circles) and $p=0.0729$ (grey squares). Bottom: $p=0.1229$ (violet diamonds), $p=0.1329$ 
(orange circles) and $p=0.1349$ (grey squares). The dashed lines indicate the best fits 
obtained at the critical points $p=p_c=0.0629(1)$ (top) and
$p=p_c=0.1329(1)$ (bottom), from which we get $\beta/\nu=0.59(1)$
(top) and $\beta/\nu=0.50(1)$ (bottom).
(Right) The average cluster size $S$ as a function of the network size $N$ for the same 
values of $p$ used for $P$. The dashed lines have slopes
$\gamma/\nu=0.24(1)$ (top) and $\gamma/\nu=0.42(1)$ (bottom). Simulations have been performed on graphs with $N=8\,192\,000$.}
\label{fig5}
\end{center}
\end{figure}
Many natural, social and man-made systems, if represented as
graphs, display common features. The most striking is a broad
distribution of the degree of the nodes, with a large majority of nodes having low degree and
a small subset of nodes having high degree. The tails of the
degree distributions are often well fitted by power laws, which lack a
characteristic scale, justifying the
name ``scale-free networks'' given to such
systems~\cite{albert02,dorogovtsev02,newman03,pastor04,boccaletti06,caldarelli07,barrat08,cohen10}. 
We indicate the exponent of the power law degree distribution with $\lambda$.
Nodes with large degree, called
``hubs'', have a fundamental role for the structure and dynamics of
networks. For instance, they ``keep'' a large portion of the system in
the same connected component, guaranteeing stability and compactness
to the structure. In particular, in random scale-free (SF) networks
with degree exponent $\lambda<3$, there are many hubs and a very small fraction of 
links (vanishing in the limit of systems of infinite size) manages to keep a macroscopic fraction of nodes of the graph
together in the same connected component. This amounts to saying that the 
percolation threshold is zero~\cite{cohen00,newman01,pastor00,dorogovtsev08,vazquez04}.

The Achlioptas process with product rule for SF networks is
implemented as follows. We start with a set of $N$ nodes and a given 
degree sequence $\{k_1, k_2, \ldots , k_N \}$. The degrees of the sequence are extracted from a 
power law distribution with exponent $\lambda$. The average degree
$\langle k\rangle$ is set equal to $5$. Every node $i$ holds $k_i$
stubs (half-links), where $k_i$ is the degree of $i$. By attaching
such stubs in pairs one builds a network with the desired power law
degree distribution with exponent $\lambda$. If stubs are
randomly connected we would have the simple configuration
model~\cite{molloy95}, which yields a random SF network. For the Achlioptas process
at each iteration two pairs of stubs are selected and the PR determines 
which pair of stubs has to be eventually joined in a link (the PR
applies as shown in Fig.~\ref{fig1}). 

The first remarkable result concerns the threshold.
For random percolation it is zero for $\lambda<3$ and non-zero for $\lambda>3$~\cite{cohen00}.
However, for explosive percolation, the threshold is non-zero already
for $\lambda>\lambda_c\sim 2.3$ (Fig.~\ref{figthres})~\cite{cho09,radicchi09}. 

In Fig.~\ref{fig5} we show the scalings at $p_c$ of $P$ and $S$ for
$\lambda=2.5$ and $2.8$. Here we see an important difference with the
cases of lattices and Erd\"os-R\'enyi random graphs, namely that 
the scaling of $P$ at $p_c$ is non-trivial, as $P$ decreases with $N$
as a power law in both cases. This is very different from what one
would find for a discontinuous transition, in which $P$ would be
approximately constant with $N$.
For $S$ we also find power law scaling (Fig.~\ref{fig5}, right
panels), as we had previously seen for lattices and Erd\"os-R\'enyi
random graphs. Overall, the transition looks like a standard
continuous transition.
\begin{figure}
\begin{center}
\includegraphics[width=0.32\textwidth]{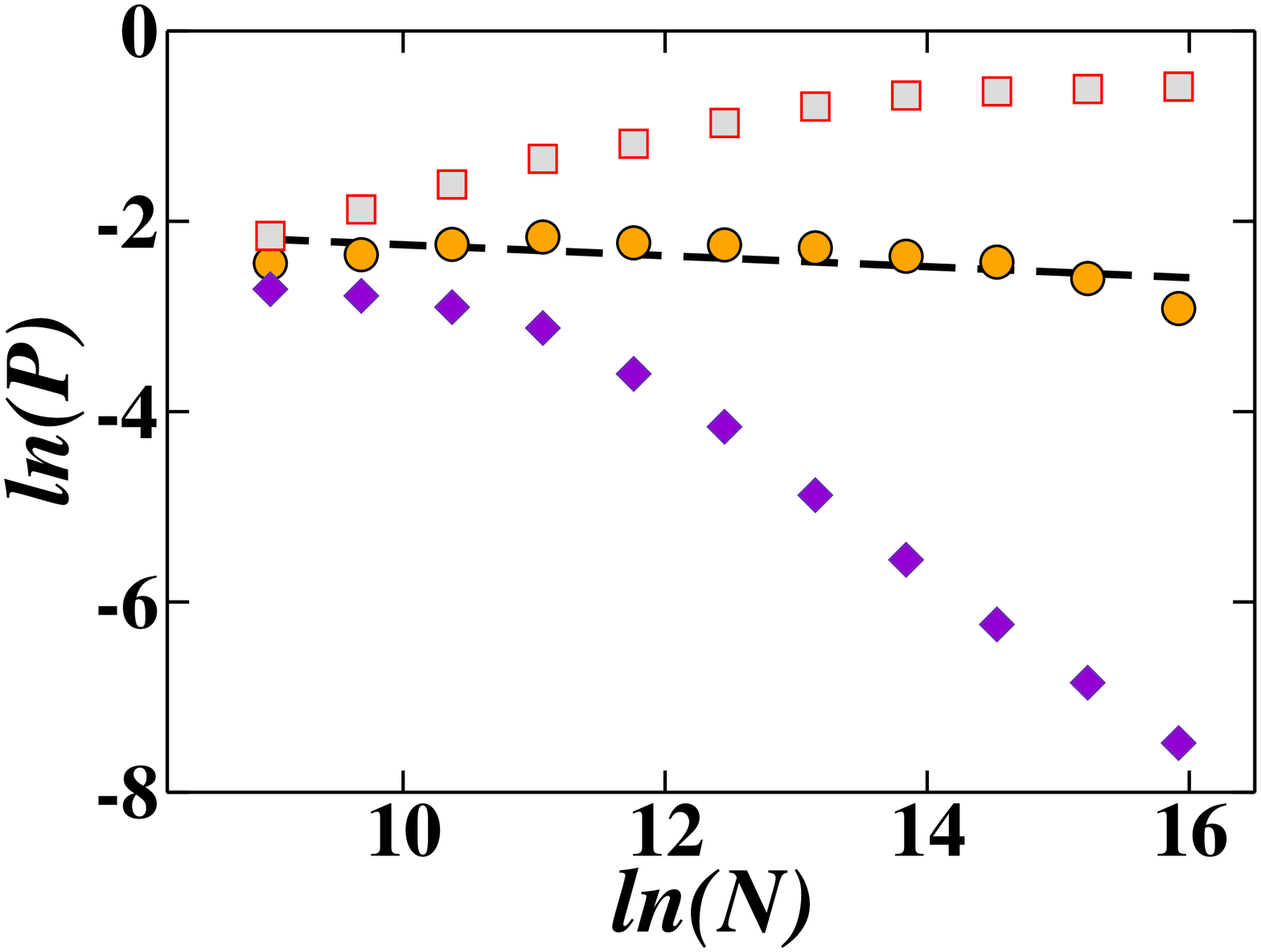}
\includegraphics[width=0.32\textwidth]{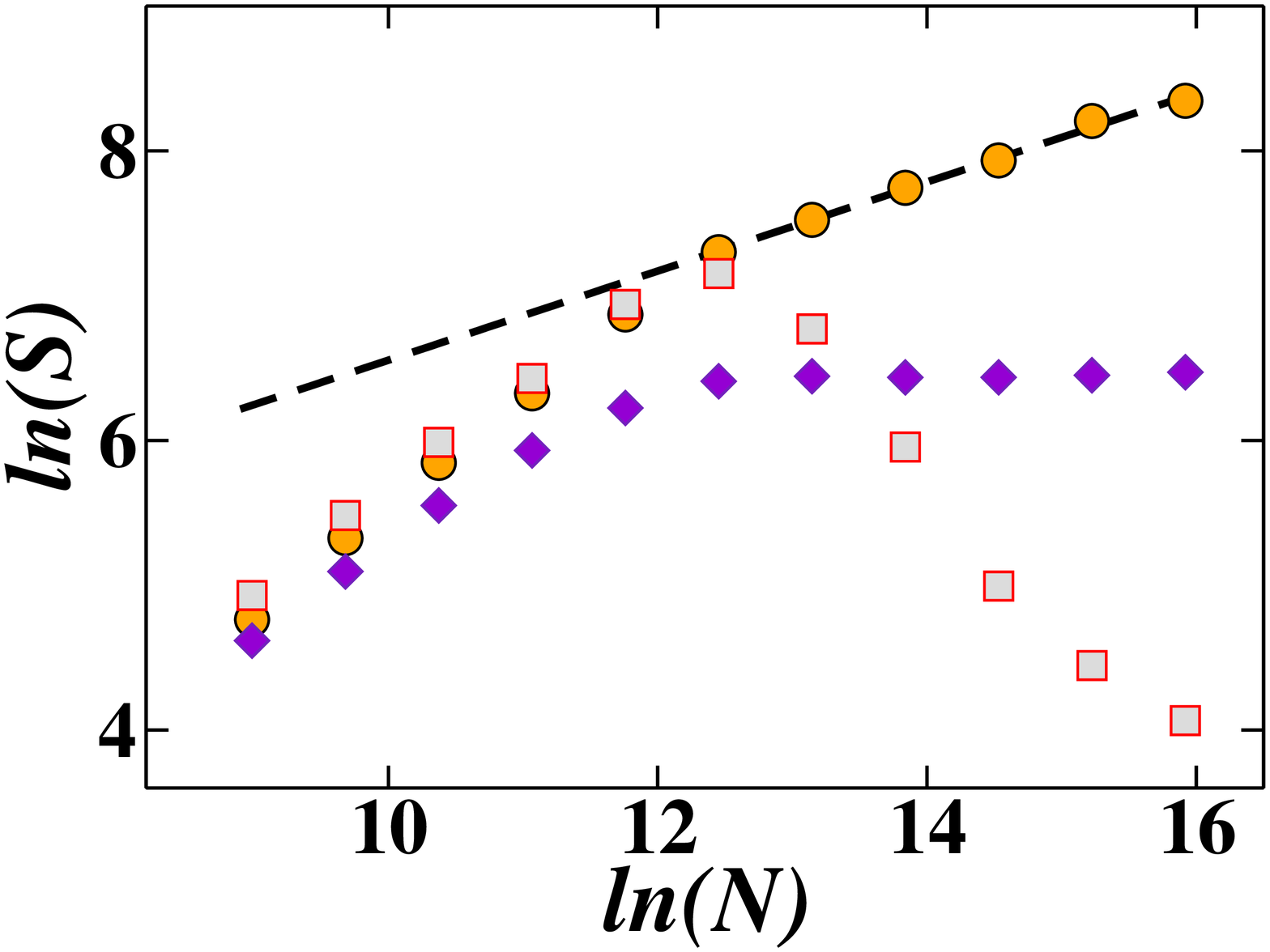}
\includegraphics[width=0.32\textwidth]{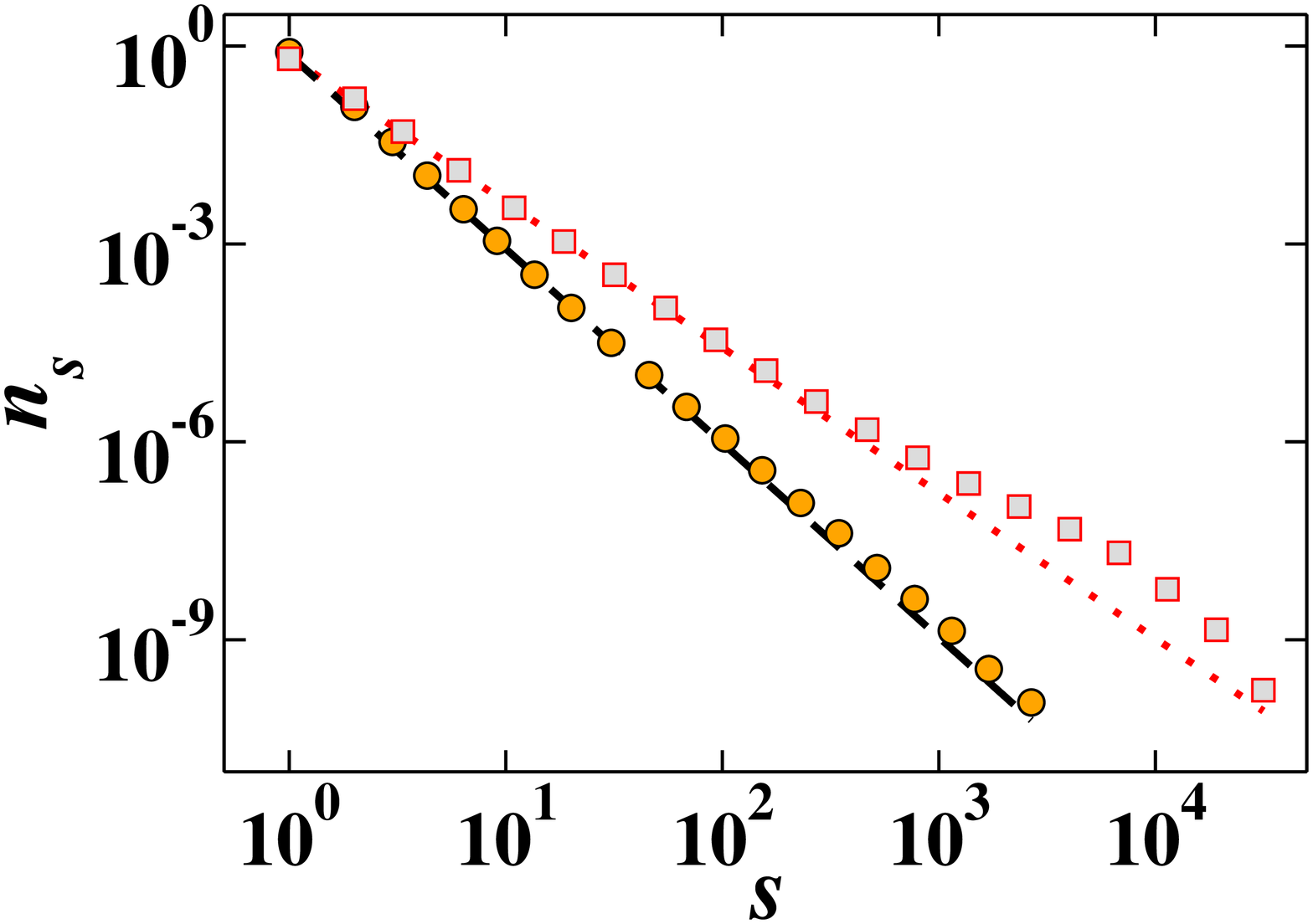}
\caption{Explosive percolation on SF networks for degree exponent
  $\lambda=3.5$.  
(Left) Percolation strength $P$  
as a function of the system size $N$ for three different values of the occupation 
probability: $p=0.2214$ (violet diamonds), $p=0.2224$ (orange circles) and $p=0.2234$ 
(grey squares). The dashed line indicates the best fit obtained at the critical point 
$p=p_c=0.2224(2)$, from which we get $\beta/\nu=-0.06(3)$. (Center) 
The average cluster size $S$ as a function of the network size $N$ for the 
same values of $p$ used for $P$. The dashed line has slope $\gamma/\nu=0.40(9)$. (Right)  
Cluster size
distributions $n_s$ measured at the critical point for PR (grey
squares) and random percolation (orange circles). The exponents are 
$\tau=2.2(1)$ (red dotted line) for PR and  $\tau=2.94(1)$ (black
dashed line) for random percolation. Simulations have been performed 
on graphs with $N=8\,192\,000$. }
\label{fig6}
\end{center}
\end{figure}
\begin{table*}
\begin{center}
\begin{tabular}{|c|c|c|c|c|c|}
\hline
System &  Model & $p_c$ & $\beta/\nu$ & $\gamma/\nu$ & $\tau$ 
\\
\hline
\hline
\multirow{2}{*}{$2d$-lattice} & RP & $0.5$ & $0.11(1)$ & $1.76(1)$ & $2.05(1)$ 
\\
\cline{2-6}
 & PR & $0.5266(2)$ & $0.07(3)$ & $1.7(1)$ & $1.9(1)$  
\\
\hline
\hline
\multirow{2}{*}{$3d$-lattice} & RP & $0.2488(3)$ & $0.48(1)$ & $2.0(1)$ & $2.20(1)$
\\
\cline{2-6}
 & PR & $0.3876(2)$ & $0.02(2)$ & $2.1(1)$ & $1.99(4)$ 
\\
\hline
\hline
\multirow{2}{*}{ER network} & RP & $0.5$ & $0.33(1)$ & $0.34(1)$ & $2.51(2)$ 
\\
\cline{2-6}
& PR & $0.8882(2)$ & $0.02(1)$ & $0.48(4)$ & $2.08(5)$ 
\\
\hline
\hline
\multirow{2}{*}{SF network $\lambda=2.5$} & RP & $0$ & $-$ & $-$ & $-$ 
\\
\cline{2-6}
& PR & $0.0629(1)$ & $0.59(1)$ & $0.24(1)$ & $2.15(2)$ 
\\
\hline
\hline
\multirow{2}{*}{SF network $\lambda=2.8$} & RP & $0$ & $-$ & $-$ & $-$ 
\\
\cline{2-6}
& PR & $0.1329(1)$ & $0.50(1)$ & $0.42(1)$ & $2.13(6)$ 
\\
\hline
\hline
\multirow{2}{*}{SF network $\lambda=3.5$} & RP & $0.078(1)$ & $0.38(1)$ & $0.15(2)$ & $2.94(1)$ 
\\
\cline{2-6}
& PR & $0.2224(2)$ & $-0.06(3)$ & $0.40(9)$ & $2.2(1)$
\\
\hline
\end{tabular}
\caption{Percolation thresholds and critical exponents for random
  percolation (RP) and Achlioptas process with product rule (PR).}
\label{table}
\end{center}
\end{table*}

For $\lambda>3$, however, the scenario changes. In Fig.~\ref{fig6} we show the results of our finite size 
scaling analysis for SF networks with exponent $\lambda=3.5$.
For random percolation on SF networks it is well known~\cite{cohen02} that $\beta/\nu=1/(\lambda-1)$ and $\gamma/\nu=(\lambda-3)/(\lambda-1)$
for $3\leq\lambda\leq 4$. For $\lambda>4$, we are in the mean field
limit and the exponents are independent of the degree exponent $\lambda$: $\beta/\nu=\gamma/\nu=1/3$. 
These values coincide with the exponents for the percolation transition of Erd\"os-R\'enyi random networks. 
SF networks are very similar to Erd\"os-R\'enyi random networks 
in the limit $\lambda\rightarrow\infty$. 
The explosive percolation transition looks as on lattices and ER random networks. 
The scaling of $P$ at $p_c$ is trivial (Fig.~\ref{fig6}, left), with $\beta/\nu=-0.06(3)$, which
is essentially zero. The power law scaling of $S$ at $p_c$ is
non-trivial (Fig.~\ref{fig6}, center), with $\gamma/\nu=0.40(9)$. The Fisher exponent 
$\tau=2.2(1)$ (Fig.~\ref{fig6}, right). 

\section{Summary}
\label{sec3}

We have found two different classes of phase transitions for explosive
percolation, that we shall call class A and B. Class A includes ER
random networks (the original system studied by Achlioptas et al.~\cite{achlioptas09}), lattices and SF networks with 
degree exponent $\lambda>3$. Here we have observed
an apparent saturation of the order parameter $P$ at $p_c$ with the
size of the system $N$, due to an exponent $\beta$ that is very small,
compatible with zero. Class B includes SF networks with
$\lambda<3$, where there seems to be no room for a discontinuous
transition, as the results of the finite size scaling analysis are
fully compatible with a standard continuous transition. 
Still, apart from the anomalous scaling behavior of the order
parameter $P$ at $p_c$ observed in class A, we found that the
other percolation variables display power
law scaling in all cases, without exceptions, just as in second-order phase transitions. The most striking
feature is the fact that the size distribution 
$n_s$ of the ``finite'' clusters at $p_c$ is 
a power law, not exponential or Gaussian as one expects from first-order phase transitions. 

Da Costa et al.~\cite{dacosta10} have recently proved that an aggregation
process in the same spirit as Achlioptas processes leads to a continuous phase transition,
characterized by an exponent $\beta$ for the order parameter $P$ that
is very small, $\beta\sim 0.0555$. They also demonstrated that the
fact that the transition is continuous for the special process they
considered enforces the same type of transition for Achlioptas
processes too, with generally different but still small values for $\beta$.
This is fully consistent with what we have found for the systems of
class A. Da Costa et al. have derived important relations for the
critical exponents. For instance, $\beta/\nu=\beta/(4\beta+1)$ and the Fisher exponent
$\tau=2+\beta/(3\beta+1)$. For $\beta\sim 0.0555$ the exponents' ratio
$\beta/\nu\sim 0.0455$. This value is compatible, within errors, with the values of
$\beta/\nu$ computed for the systems of class A (see Table 1). So, the
apparent saturation of the order parameter $P$ with the system size
$N$ could be indeed due to the smallness of the exponent
$\beta$. Moreover, the Fisher exponent $\tau$ for every transition we
have investigated is very close to $2$. From the relation 
$\tau=2+\beta/(3\beta+1)< 2+\beta$, and the fact that $\beta$ is small, we see
that $\tau$ will always be very close to $2$, just as we found. The
difference between $\tau$ and $2$ is hard to determine numerically. Likewise, the values of $\tau$ derived from different
Achlioptas processes can be hardly distinguished from each other.
We remark that this holds for lattices, Erd\"os-R\'enyi graphs and SF
networks with degree exponent $\lambda>3$, while the
explosive percolation transition on SF networks with $\lambda<3$
appears as something very different ($\beta$ is not small here), and deserves further investigations in the future.

We have also found that there is a non-zero explosive percolation threshold for SF networks 
for $\lambda>\lambda_c\sim 2.3$ (Fig.~\ref{figthres}), while the
threshold for random percolation is zero as long as $\lambda \leq 3$.
Cho {\it et al.}~\cite{cho09} suggested that this is due to the non-random addition
of links during the Achlioptas process, because of which the degree distribution of the system during the growth 
deviates from that imposed by construction, which will be eventually
reached at the end of the process. This has been indeed verified
numerically~\cite{cho09,radicchi10}.

\ack
S. F. gratefully acknowledges ICTeCollective, grant number 238597 of the European Commission.

\end{document}